\documentclass[conference]{IEEEtran}
\IEEEoverridecommandlockouts
\usepackage{amsmath,amssymb,amsfonts}
\usepackage{algorithmic}
\usepackage{graphicx}
\usepackage{textcomp}
\usepackage{caption}
\usepackage{xcolor}
\usepackage{subfigure}
\usepackage{titlesec}
\def\BibTeX{{\rm B\kern-.05em{\sc i\kern-.025em b}\kern-.08em
    T\kern-.1667em\lower.7ex\hbox{E}\kern-.125emX}}
\usepackage{hyperref}
\usepackage{biblatex}
\begin{document}

\title{TDCGL: Two-Level Debiased Contrastive Graph Learning for Recommendation
\thanks{\textsuperscript{*} The authors contribute equally to this paper.\\}
}

\author{\IEEEauthorblockN{1\textsuperscript{st} Yubo Gao*}
\IEEEauthorblockA{
\textit{North University of China}\\
Taiyuan, China \\
gyb1447437905@gmail.com}
\and
\IEEEauthorblockN{2\textsuperscript{nd} Haotian Wu*}
\IEEEauthorblockA{
\textit{Beijing Jiaotong University}\\
Beijing, China \\
wu\_haotian@bjtu.edu.cn}
}
\maketitle

\begin{abstract}
As a milestone research combining recommender systems and knowledge graphs (KG), Knowledge Graph Attention Network (KGAT) has achieved great success in the field of recommender systems by proposing a new approach that implements explicit end-to-end modeling of higher-order relationships in a graph neural network framework to provide better item-assisted information recommendations. A series of methods following KGAT provide more solutions for KG-based recommendations. However, over-reliance on high-quality knowledge graphs is a bottleneck for such methods. Specifically, the long-tailed distribution of entities of KG and noise issues in the real world will make item-entity dependent relations deviate from reflecting true characteristics and significantly harm the performance of modeling user preference. Contrastive learning, as a novel method that is employed for data augmentation and denoising, provides inspiration to fill this research gap. However, the mainstream work only focuses on the long-tail properties of the number of items clicked, while ignoring that the long-tail properties of total number of clicks per user may also affect the performance of the recommendation model. Therefore, to tackle these problems, motivated by the Debiased Contrastive Learning of Unsupervised Sentence Representations (DCLR), we propose Two-Level Debiased Contrastive Graph Learning (TDCGL) model. Specifically, we design the Two-Level Debiased Contrastive Learning (TDCL) and deploy it in the KG, which is conducted not only on User-Item pairs but also on User-User pairs for modeling higher-order relations. Also, to reduce the bias caused by random sampling in contrastive learning, with the exception of the negative samples obtained by random sampling, we add a noise-based generation of negation to ensure spatial uniformity. Considerable experiments on open-source datasets demonstrate that our method has excellent anti-noise capability and significantly outperforms state-of-the-art baselines. In addition, ablation studies about the necessity for each level of TDCL are conducted.
\end{abstract}


\vspace{-0.4cm}
\section{Introduction}
\vspace{-0.2cm}
Recommender systems, which aim to predict users' interests and filter, prioritize and recommend items that users prefer, are efficient solvers for the information overload phenomenon \cite{wu2022mncm,wu2023mt}. As a result, recommender systems have become a promising research direction in the field of artificial intelligence in recent years \cite{zhang2022ctnocvr}. The collaborative filtering (CF) \cite{rendle2020neural} framework is an effective solution for predicting user preferences, which has evolved from matrix factorization (MF) to the latent user and item embedding projection techniques based on various neural networks (e.g. Autorec \cite{sedhain2015autorec}, ACF \cite{chen2017attentive}, LightGCN \cite{he2020lightgcn}), and its modeling capability for complex User-Item interaction patterns has been improved. 

In the field of recommender systems, in order to make the recommendation results more accurate, it is necessary not only to consider the relationship between User-Items but also to introduce side information to enrich the information between User-Item. However, the CF framework cannot model side information such as item attributes, user profiles, and contexts, thereby performing poorly when there are few User-Item interactions \cite{bordes2013translating}. To tackle this challenge, a common current approach is to transform the side information along with the user ID and item ID into a generic feature vector and feed them into a supervised learning (SL) model to predict scores \cite{wang2019kgat,he2017neural}. The Knowledge Graph (KG) provides an effective solution to discard the independent interactions assumption \cite{wang2019kgat}. As a heterogeneous graph structure, the nodes of KG function as entities, the edges represent the relationships among entities, and items and their attributes can be mapped to KG to reflect the interrelationships between items \cite{zhang2016collaborative}. In addition, user and user-side auxiliary information can also be incorporated into the KG, which can facilitate capturing user preferences for items more effectively \cite{guo2020survey}. In KG, users and items are connected by different potential relations, containing higher-order relations as well (e.g. long-range connectivities), which facilitates more accurate recommendations \cite{wang2019kgat}. 

Prior studies on introducing KG as auxiliary information into recommender systems are divided into three main groups: Embedding\-based methods, Path\-based Methods, and Unified methods \cite{guo2020survey}. Embedding\-based methods \cite{bordes2013translating,lin2015learning,zhang2018learning} usually directly use the Knowledge Graph Embedding (KGE) algorithm to encode KGs as low-rank embeddings to exploit the information in KGs to enrich the item and user representations. The Path-based methods \cite{shi2015semantic,sun2018recurrent,ma2019jointly} construct User-Item graphs and utilize the connectivity similarities between entities to make recommendations. 
However, both embedding\-based and path\-based approaches can only utilize information from one graph aspect. To fully utilize the information of knowledge graphs, unified methods that combine semantic representations of entities and relations and connectivity information have been proposed (e.g. RippleNet \cite{wang2018ripplenet}, AKUPM \cite{tang2019akupm}, RCoLM \cite{li2019unifying}, KGCN \cite{wang2019knowledge}, KGAT  \cite{wang2019kgat}). These knowledge graph-based recommendation methods suffer from a common shortcoming: they are over\-reliant on the quality of the knowledge graph, which is reflected in the following limitation: High-degree nodes exert a larger impact on the representation learning, deteriorating the recommendations of low-degree (long-tail) items. What's more, we find that the long tail problem is not only at the item level but also at the user level, which may significantly affect the model's performance and is ignored by prior studies. 

In this paper, we start with the famous methods in KG-based recommendation: Knowledge Graph Attention Network for Recommendation (KGAT) \cite{wang2019kgat}, which links User-Item instances together through the attributes between user and item, and fuses User-Item and knowledge graph together to form a new network structure, and extracts higher-order linking paths from this network structure to express the nodes in the network. Contrastive learning \cite{he2020momentum,chen2020simple} provides the initial inspiration for our research to allivate the negative impact of long-tail issues, which achieves success in several studies \cite{wu2021sgl}. However, existing studies only consider the long-tail property of items while ignoring the fact that there is also a long-tail property of users. Specifically, some users always interact with items significantly more often than others. To this end, we propose a novel Two-Level Debiased Contrastive Learning (TDCL) and deploy it in KG for the recommendation, namely Two-Level Debiased Contrastive Graph Learning (TDCGL) model. TDCL is mainly motivated by Debiased Contrastive Learning of Unsupervised Sentence Representations (DCLR) \cite{zhou2022debiased}, a framework for debiased sentence representation contrastive learning for the problem of wrong-negative cases and anisotropy. In TDCL, we design two sorts of negative samples at the User-User level and User-Item level, which are randomly drawn from in-batch and generated based on white noise distribution (Noise-based samples) respectively. Specifically, considering that the distribution of vectors randomly drawn from in-batch is usually not ideal, the introduction of noised-based samples makes the vectors evenly distributed in space. In addition, we penalize false negatives through the instance weighting method to mitigate the impact of false negative 
samples on the performance of contrastive learning at the User-User level and User-Item level. What's more, we introduced the projection head following the SimCLR \cite{chen2020simple} to preserve the original semantics through the nonlinear activation layer. 

On the open-source recommendation dataset, we conducted comparison experiments with the state-of-art methods in several recommender systems to verify the effectiveness of TDCGL. The experimental results show that our proposed TDCGL  effectively mitigates the negative impact of the long-tail distribution of entities in the knowledge graph, and thus TDCGL significantly outperforms baselines. Additionally, ablation studies were designed to demonstrate the necessity of introducing two levels of TDCL. 
\vspace{-0.4cm}
\section{Approach}
\begin{figure}[htp]
    \centering
    \includegraphics[width=0.45\textwidth]{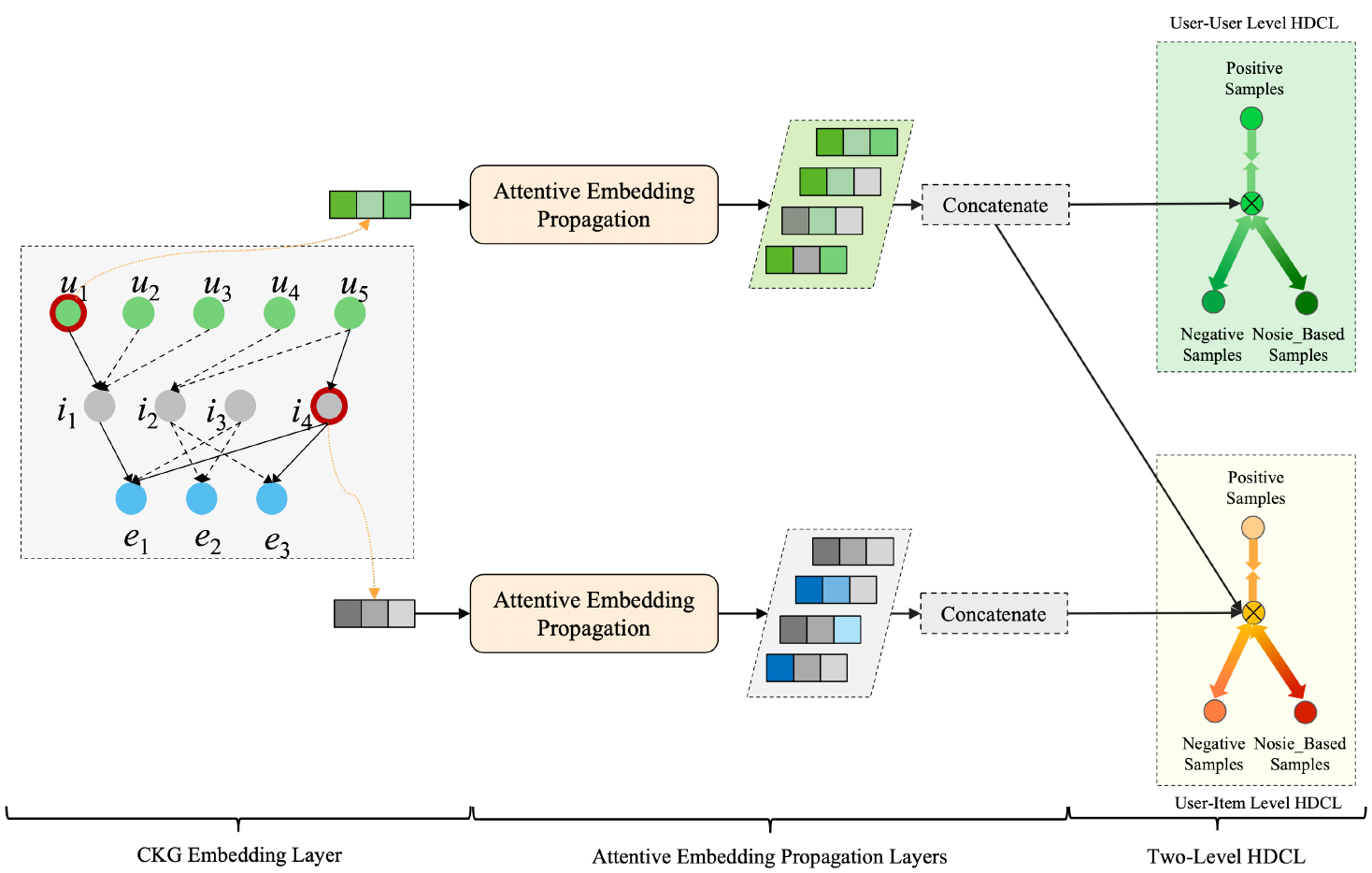}
    \caption{The Overall Framework of Two-Level Debiased Contrastive Graph Learning (TDCGL)}
    \label{fig3}
\end{figure}

\vspace{-0.2cm}
In this section, we will introduce the details of our model. For the given user and the product data that the user has clicked, we predict the product that the user will click. Concretely, first, we get positive samples and negative samples through the data, then perform two-level debiased contrastive learning on the User-User and User-Item levels respectively. Finally, add $KG-loss$ through the knowledge graph. Figure ~\ref{fig3} presents the TDCGL in a nutshell.
\vspace{-0.3cm}

\subsection{Bayesian Personalized Ranking Loss}
\vspace{-0.2cm}
We employ the BPR loss \cite{sun2018recurrent} to optimize the recommendation model, which assumes that the observed interactions indicating more user preferences, should be assigned higher prediction values than unobserved ones:
\begin{equation}
    \mathcal{L}_{\mathrm{CF}}=\sum_{(u, i, j) \in D}-\ln \sigma(\hat{y}(u, i)-\hat{y}(u, j))
\end{equation}
where $D$ denotes the training dataset, $u$ represents a distinct user, $i$ denotes an item that has a interaction with $u$ and $j$ denotes an item that does not have any interaction with $u$. $\sigma(\cdot)$ is the sigmoid function.

\subsection{Two-level Debiased Contrastive Learning}
Given a negative embedding $e^-$ and the embedding $e$, we utilize the complementary model to produce the weight as:
\begin{equation}
    \alpha=\left\{\begin{array}{l}
0, \operatorname{sim}_{C}\left(e, e^{-}\right) \geq \phi \\
1, \operatorname{sim}_{C}\left(e, e^{-}\right)<\phi
\end{array}\right.
\end{equation}
where $\phi$ is a hyper-parameter of the instance weighting threshold, and ${sim}_{C}\left(e, e^{-}\right)$ is the similarity score evaluated by the complementary model.

In terms of two-level debiased contrastive learning, with the User-User level and User-Item level. The User-Item in-batch loss is denoted as the contrastive loss between the in-batch positive pairs $(u, i)$ and the in-batch negative pairs $(u, j)$ as:
\begin{equation}
    \hat{\mathcal{L}_{\mathrm{ui}} } = -\log\frac{e^{sim(u, i)/\tau}}{\sum_{j \in J} \alpha \times e^{sim(u, j)/\tau}}
\end{equation}
where $\tau$ is a temperature hyper-parameter and $sim$ is the similarity score evaluated by the model, $J$ is a set of in-batch negatives. Assuming that $u^+$ is $u$ itself and $o$ is all users except $u^+$, the User-User in-batch loss is denoted as: 
\vspace{-0.2cm}
\begin{equation}
    \hat{\mathcal{L}_{\mathrm{uu}} } = -\log\frac{e^{sim(u, u^+)/\tau}}{\sum_{o \in O} \alpha \times e^{sim(u, o)/\tau}}
\end{equation}
Inspired by DCLR \cite{zhou2022debiased}, we also propose a novel two-level noised loss to mitigate the false negative samples issues, where $i^-$ denotes negatives based on random Gaussian noises.
The User-Item noised loss is denoted as the contrastive loss between the positive pairs $(u, i)$ and the noised negative pairs $(u, i^-)$ as:
\begin{equation}
    \mathcal{L}_{\mathrm{ui} }^{-} = -\log\frac{e^{sim(u, i)/\tau}}{\sum_{i^- \in I^-} e^{sim(u, i^-)/\tau}}
\end{equation}
where $I^-$ is a set of noised negatives. Assuming $u^-$ is a set of the noised negatives of $u$, the User-User noised loss is denoted as:

\begin{equation}
    \mathcal{L}_{\mathrm{uu} }^{-} = -\log\frac{e^{sim(u, u^+)/\tau}}{\sum_{u^- \in U^-} e^{sim(u, u^-)/\tau}}
\end{equation}
\vspace{-0.3cm}
\subsection{Knowledge Graph Loss}
Knowledge graph embedding provides an effective means to parameterize entities and relationships in vector representations, while preserving the underlying graph structure. In this study, we apply the TransR\cite{lin2015learning} approach, a widely accepted methodology, in the context of collaborative knowledge graphs (CKGs). To elaborate, TransR detects and encapsulates each entity and relationship by optimizing the translation principle: 
$\mathbf{e}_{h}^{r}+\mathbf{e}_{r} \approx \mathbf{e}_{t}^{r}$
if a triplet $(h,r,t)$ exists in the graph. In this context, $\mathbf{e}_{h}, \mathbf{e}_{t} \in \mathbb{R}^{d}$ and $\mathbb{R}^{k}$ respectively serve as the embeddings for the entities $h$ and $t$ along with $r$, while $e_{h}^{r}$, $e_{t}^{k}$ represent the projected manifestations of $e_{h}$ and $e_{t}$ within the spatial boundaries of the relation $r$.
Consequently, for a given triplet $(h,r,t)$, its plausibility score, often called the energy score, is expressed as:
\vspace{-0.2cm}
\begin{equation}
    g(h, r, t)=\left\|\mathbf{W}_{r} \mathbf{e}_{h}+\mathbf{e}_{r}-\mathbf{W}_{r} \mathbf{e}_{t}\right\|_{2}^{2}
\end{equation}

Here, $\mathbf{W}_{r} \in \mathbb{R}^{k \times d}$ acts as the transformation matrix for the relation $r$. It performs the transformation of entities from the $d$-dimensional entity space to the $k$-dimensional relation space. In particular, a lower score within $g(h,r,t)$ conveys a higher probability that the triplet is indeed true, and conversely, a higher score signifies the opposite. The training regimen of TransR meticulously considers the relative ordering between valid triplets and their defective counterparts. This discrimination is actively encouraged through the employment of a pairwise ranking loss function, as represented by:
\vspace{-0.2cm}
\begin{equation}
    \mathcal{L}_{\mathrm{KG}}=\sum_{\left(h, r, t, t^{\prime}\right) \in \mathcal{T}}-\ln \sigma\left(g\left(h, r, t^{\prime}\right)-g(h, r, t)\right)
\end{equation}

where $\mathcal{T}=\left\{\left(h, r, t, t^{\prime}\right) \mid(h, r, t) \in \mathcal{G},\left(h, r, t^{\prime}\right) \notin \mathcal{G}\right\}$, and $(h,r,t^{\prime})$ being artificially constructed to be defective by randomly substituting one entity within a valid triplet; $\theta(\cdot)$ is the sigmoid function. This layer operates at the granularity of triplets, functioning as a regularizer that injects direct interconnections into the representations. As a result, it significantly enhances the model's representation capabilities.

We calculate the losses of $\mathcal{L}$ and $\mathcal{L}_{KG}$ separately, and the loss of $\mathcal{L}$ is:
\begin{equation}
    	\mathcal{L}=\mathcal{L}_{\mathrm{CF}}+\hat{\mathcal{L}_{\mathrm{ui}} }+\hat{\mathcal{L}_{\mathrm{uu}} }+\mathcal{L}_{\mathrm{ui} }^{-}+\mathcal{L}_{\mathrm{uu} }^{-}+\lambda\|\theta\|_{2}^{2}
\end{equation}

The final loss is:
\begin{equation}
    \mathcal{L}_\mathrm{final} = \mathcal{L} + \mathcal{L}_\mathrm{KG}
\end{equation}

\section{Experiment}
\vspace{-0.3cm}
\subsection{Dataset}
In addition to user-item interactions, it is imperative to establish item knowledge for each dataset. For ML-1M and Amazon-book (we retain users and items with at least twenty-five interactions.) datasets, the statistical details of the experimented datasets are presented in Table ~\ref{table1}.

\vspace{-0.2cm}
\begin{table}[htp]
\centering
\caption{Statistics of experimented datasets}
\label{table1}
\begin{tabular}{c|cc}
\hline
Stats         & \multicolumn{1}{c|}{ML-1M}  & Amazon-Book \\ \hline
\#Users       & \multicolumn{1}{c|}{6040}   & 16330       \\ 
\#Items       & \multicolumn{1}{c|}{3629}   & 18413       \\
\#Interaction & \multicolumn{1}{c|}{836478} & 1033067     \\ \hline
              & \multicolumn{2}{c}{Knowledge Graph}       \\ \hline
\#Entities    & \multicolumn{1}{c|}{79388}  & 200601      \\
\#Relations   & \multicolumn{1}{c|}{51}     & 22          \\
\#Triplets    & \multicolumn{1}{c|}{385923} & 522475      \\ \hline
\end{tabular}
\end{table}
\vspace{-0.4cm}

\subsection{Experimental Settings}
\vspace{-0.1cm}
\subsubsection{Evaluation Metrics.}
We treat all items with which the user has not interacted as negative items for each user in the test set. Subsequently, each methodology yields user-specific preference scores encompassing the entire item pool, excluding those affirmed in the training set. To gauge the performance of top-K recommendation and preference ranking, we employ two established assessment frameworks: NDCG@K and Recall@K. The default setting prescribes K to be 10, and our findings report aggregate metrics across the entirety of test users.

\subsubsection{Baseline Methods.}
We compare TDCGL with competitive lines of recommender systems and unsupervised sentence representation learning methods:
\begin{enumerate}
    \item{\bf LightGCN \cite{he2020lightgcn}}. This represents a cutting-edge recommendation approach based on Graph Convolutional Networks (GCNs). 
    \item{\bf SGL \cite{wu2021sgl}}. Employing an augmented structure-based self-supervised signaling system, SGL demonstrates superior performance within the context of graph-based Collaborative Filtering (CF) frameworks.
    \item {\bf KGCN \cite{wang2019knowledge}}. It aims to improve collaborative filtering recommender systems by using knowledge graphs to capture item relatedness, mitigating sparsity and cold start problems.
    \item {\bf RippleNet\cite{wang2018ripplenet}}. RippleNet enhances click-through rate prediction in knowledge-graph-aware recommendation by combining embedding and path-based methods. 
    \item {\bf KGAT \cite{wang2019kgat}}. KGAT improves the accuracy and interpretability of recommendations by modeling high-order relationships in collaborative knowledge graphs using item-side information.
    \item {\bf KGAT+DCLR}. It adds de-biased contrastive learning to KGAT to verify the effect of randomly generated negative examples.
    \item {\bf KGATS+CTS}. It adds vanilla contrastive learning to KGAT while modeling the relationship between pairs of User-Item only.
    \item {\bf KGATB-CTS}. It adds vanilla contrastive learning to KGAT while modeling the relationship between User-User and User-Item at the same time.
\end{enumerate}

\subsubsection{Parameter Settings.}
The embedding size is fixed to 64 for all models.
All of the compared baselines are evaluated based on the unified recommendation library RecBole \cite{zhao2021recbole}. In particular, we fix the embedding dimensionality as 64 for all methods and conduct the model optimization with a learning rate of 0.001 and batch size of 4096. For knowledge-aware recommendation models, the number of context hops and memory size are set as 2 and 8, respectively. 
In our TDCGL, we search the temperature parameter $\tau$ in the range of
{0.01,0.1,0.5,1.0} with an increment of 1.0. Moreover, an early stopping strategy is conducted, premature stopping if Recall\@10 on the validation set does not increase for 50 successive epochs. 
\vspace{-0.2cm}
\subsection{Experimental Results}
We show the overall performance evaluation of all methods in Table ~\ref{table2}. From the results, TDCGL performs better than the other baselines in all cases. The diversity of the evaluated datasets varies depending on the sparsity, knowledge graph features, and recommendation scenarios. The excellent results demonstrate the universality and importance of the knowledge graph, and the excellent results demonstrate the generality and flexibility of our TDCGL framework. The performance of KGAT and KGCN is weaker than that of SGL, KGAT+DCLR, KGAT+CTS, KGAT-CTS and TDCGL, indicating that the introduction of contrastive learning effectively mitigates the negative impact of the long-tail distribution problem on the KG-based recommendation model.
\vspace{-0.6cm}
\begin{table}[htp]
\centering
\setlength{\belowcaptionskip}{10pt}
\caption{Performance comparison of baselines and TDCGL on ML-1M and Amazon-Book}
\label{table2}
\begin{tabular}{c|cc|cc}
\hline
Datasets        & \multicolumn{2}{c|}{ML-1M}        & \multicolumn{2}{c}{Amazon-Book} \\ \hline
Methods         & NDCG         & Recall       & NDCG        & Recall      \\ \hline
LightGCN        & 0.2052              & 0.1461              & 0.0623              & 0.0728              \\
KGCN            & 0.1863              & 0.1276              & 0.0331              & 0.039              \\
KGAT            & 0.1938          & 0.1554          & 0.0523              & 0.0525              \\
RippleNet       & 0.1438              & 0.1067              & 0.0244              & 0.0281              \\
SGL             & 0.2131          & 0.1557          & 0.0695              & 0.0811              \\ \hline
KGAT+DCLR   &0.1974 & 0.1592 & 0.0543 & 0.0647 \\
KGATS+CTS       & 0.1900          & 0.1509          & 0.0426              & 0.0509              \\
KGATB-CTS       & 0.1956          & 0.1581          & 0.0709              & 0.0825              \\ \hline
\textbf{TDCGL} & \textbf{0.2134} & \textbf{0.1746} & \textbf{0.071}     & \textbf{0.0828}     \\ \hline
\end{tabular}
\end{table}
\vspace{-0.2cm}

In summary, TDCGL enhancements can be summarized in three ways.
Benefiting from our knowledge graph comparison learning, TDCGL can remove entity dependencies and capture accurate item semantics. TDCGL is able to improve the performance of contrastive learning by reducing the bias caused by random sampling in contrastive learning. TDCGL learns by performing a comparison between User-User and User-Item to obtain efficient and robust feature representations which  contain high-order relationship information.

\begin{figure}[htbp]
	\centering
	\begin{minipage}{0.49\linewidth}
		\centering
		\includegraphics[width=\linewidth]{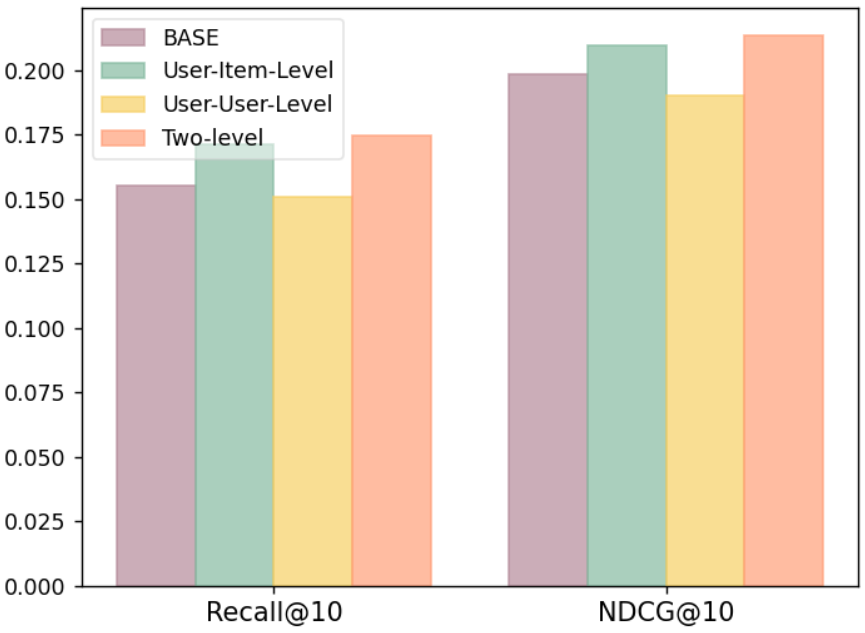}
		\caption{The necessity of each levels of TDCL}
		\label{ablation}
	\end{minipage}
	\begin{minipage}{0.49\linewidth}
		\centering
		\includegraphics[width=\linewidth]{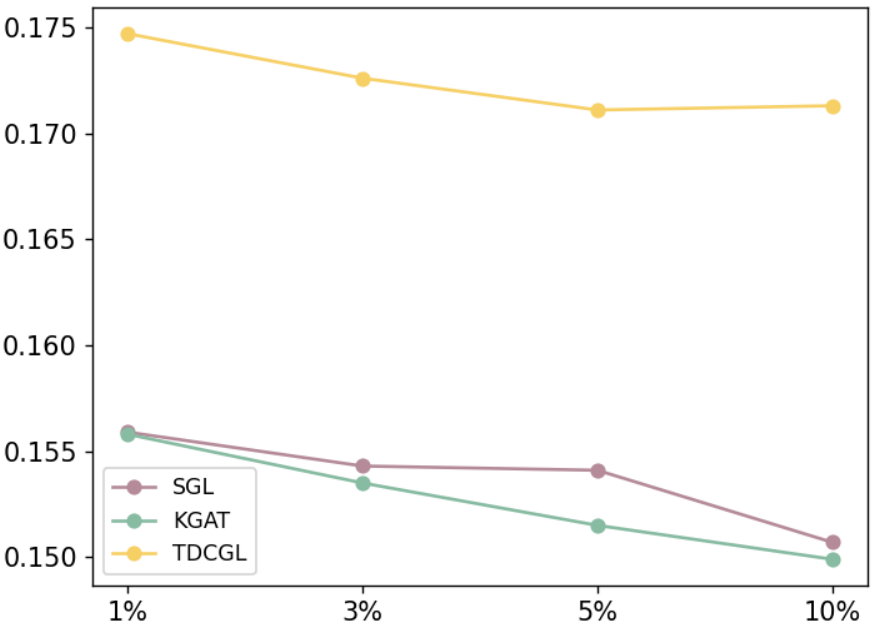}
		\caption{Comparison of anti-noise performance}
    	\label{Noise}
	\end{minipage}
\end{figure}

To verify the necessity of introducing TDCL, we set two single-level TDCL varients (User-Item and User-User) and compared them with BASE (KGAT) and Two-Level TDCGL, and the experimental results are shown in Fig. ~\ref{ablation}. 
In addition, to demonstrate the effectiveness of proposed TDCGL under different levels of noise. We randomly drop the nodes in the graph on ML-1M dataset with different rate, the experimental results of which are shown in the Fig. ~\ref{Noise}.

TDCL with User-User Level and User-Item Level outperforms TDCL with either level alone. This illustrates that this two-level contrastive learning paradigm is more suitable for modeling KG-based recommendation tasks. In addition, we observe that introducing TDCL at User-Item Level improves the performance of BASE more significantly than introducing TDCL at User-User Level, which indicates that modeling the User-Item relationship is still the ideal paradigm for KG-based recommendation methods. Although the introduction of User-User Level TDCL alone does not achieve the desired results, the experimental results combined with Two-Level TDCL demonstrate that User-User Level TDCL introduces more information about the User-User relationship, which may be neglected by User-Item Level TDCL. User-User Level TDCL enriches the feature representation to a certain extent. In addition, TDCGL significantly outperforms KGAT and SGL at different levels of noise. With the increase of dropped nodes, the performance of TDCGL, KGAT, and SGL all degrade to varying degrees
\vspace{-0.4cm}
\vspace{-0.4cm}

\vspace{-0.1cm}
\section{Conclusion and Future Work}
In this work, our proposed TDCGL framework mitigates the noisy interaction effects of KGAT due to domain aggregation. And we model higher-order relations at both levels, also penalizing negative examples to ensure spatial uniformity. Extensive experiments
on several real-world datasets have demonstrated the superiority of TDCGL as compared to various state-of-the-art methods. In addition, we also explored the effect of different layers of the projection head-on performance and the effect of different activation functions on the projection head. Furthermore, numerous experiments have proven that TDCGL has outstanding noise immunity
This work makes a preliminary attempt to explore the potential of contrastive learning and graph neural networks in recommender systems. It may be an exciting direction to get good and interpretable negative examples. Also, the problem of data noise for graph neural networks is a direction worth exploring.

\printbibliography

\end{document}